**This is a preprint of the following chapter**
Margareta Segerståhl and Matti Jalasvuori
System-Evolutionary View of RNA Presence in Living Cells, published in Integrative Theory of Evolution - Aspects and Insights, edited by Guenther Witzany, 2026, Springer, reproduced with permission of Springer Nature Switzerland AG, final authenticated version is available online at: https://doi.org/10.1007/978-3-032-24295-2_15

# System-Evolutionary View of RNA Presence in Living Cells

**Margareta Segerståhl[1, 2,*] and Matti Jalasvuori[2, 3]**

[*]Corresponding author (M.S.) margareta.segerstahl@gmail.com
[1]Current contact information: Loviisa, Finland
[2]Department of Biological and Environmental Science, Nanoscience Center, University of Jyväskylä, Jyväskylä, Finland
[3]PrecisionPhage Oy, Jyväskylä, Finland

**Abstract**
RNA predates cells but today it "lives" inside them. Most current RNA is short-lived, folded, and often edited from its chromosomal origins but it governs its own phenotypes. Our guiding question for this article was: How did RNA adjust its life when the RNA world era came to an end? RNA has its practical execution side in the whole-cell systems setup serving critical functions in metabolism, compartmentalization, and in the RNA–DNA–Protein network, coupling information handling and different events of macromolecule production into a self-sustaining loop. Aligning with current views of cooperation we find that RNA life is distributed across a population of information-bearing entities persisting by compatibility, with DNA serving as a long-term memory and proteins offering the catalytic reach. We suggest that the reproduction–survival (R/S) axis, relevant to higher organisms, is also manifested by RNA in many forms, including RNA-primers, mRNAs, tRNAs, and rRNAs. Each of these have contrasting, highly adapted molecular phenotypes. Ultimately, we wish to emphasize the evolutionary narrative where RNA did not "lose" in exchange for DNA genes or otherwise. Ultimately, RNA just changed strategy and in its various forms has an active and critically important presence in all cells at all times.

## 1.1 Background and Motivation

RNA is a key substance of biological life that dates back to ancient biochemical systems, predating the evolution of the last universal common ancestor (LUCA), or a community of common ancestors, from which extant cellular life descend from (Koonin and Martin 2005). The initial understanding of cellular RNA molecules was mainly based on the significant universal contribution of messenger RNA (mRNA), transfer RNA (tRNA) and ribosomal RNA (rRNA) molecules to the mechanisms of protein-coding gene expression. What was realized is that functional RNA sequence is, in general, synonymous to (functional) life. This on the other hand stems the question how much of life is linked to RNA. Here, we argue that ultimately RNA just changed its original strategy and life, as we know it, is derived from this strategic adaptation.

The past few decades of experimental work have led to the remarkable realization that RNA presence in cells is much more diverse and complex than previously thought. In addition to tRNA, mRNA, and rRNA there are also many additional types of RNAs with diverse functional roles, including regulators of gene expression (reviewed, e.g., by Morris and Mattick 2014). The mechanisms of action vary. Not only can non-coding (nc) RNAs modify other RNA transcripts and regulate the dynamics of many different steps of protein-coding gene expression, but they can also influence what genes can be active in the first place by introducing epigenetic modifications, such as methylations or adenylations, to DNA nucleotides, that can change expression patterns without changing the actual genomic sequence of DNA strands. Many ncRNAs can regulate cellular development and organismal differentiation and act as causal agents of phenotypic change that may even be inherited by progeny, meaning, that ncRNAs can introduce evolutionary novelty to the genetic information content of cells (reviewed, e.g., in Fitz-James and Cavalli 2022; Sánchez-Romero and Casadesús 2020) and together with other distinct mechanisms, such as horizontal gene transfer (HGT), transposable elements, and gene duplication (e.g., Schaack et al. 2010; Soucy et al. 2015), give rise to new phenotypic lineages influencing evolutionary trajectories.

The realization that RNA strands are significant functional components of cell biology regulating and generating phenotypic diversity has had an impact on the evolutionary theory. It has even been argued, that in its current state evolutionary theory is unable to accommodate and adequately explain the adaptive significance behind the observation that natural living cells have many kinds of non-random cell biological mechanisms for generating phenotypic diversity, and that some revision or update of the theory is required (see, e.g., Shan 2024; Witzany 2021, for general accounts). In practical observational terms, various studies have, of course, shown phenotypic plasticity in organisms arising from various environmental or other factors. Nevertheless, this issue is a complex problem that touches many areas of biological study where phenotype ultimately links to RNA, in one way or another and poses a *challenge for the long-term dogma, that diversity (i.e., material for natural selection) is provided by the emergence of new hereditary phenotypes by random mutation* of *DNA nucleotides.* Resolving it may well require making major changes in some previously preferred paradigmatic viewpoints, and it might be especially important to find new methods for holistic management and integration of information between different fields, especially between molecular cell biology and evolutionary biology (Shan 2024). Various perspectives, such as organismal or genetic level, can be taken to better understand the reality behind all this. Here, we, in a sense, interpolate viewpoints and provide an account from the perspective of RNA.

RNA is a key part of cellular life. Not only does it have the above-mentioned functions but with increasing knowledge of cellular long non-coding RNAs (lncRNAs), the research is already progressing towards potential applications in the medical studies (Mattick et al. 2023). Having a sound theoretical understanding of the evolutionary foundation of the numerous different kinds of non-coding RNAs is not merely an academical issue, but one that connects us with the very basis of our ability to understand the cell and its behavior—as an individual unit and as a part of the whole. In an attempt to better understand the many roles that RNA has in cellular life, we focus in this article on the topic of combining evolutionary and cell biological viewpoints in a new way, using a novel system-level model of cell biology as a framework (Segerståhl & Segerståhl 2022). This model was initially developed to describe the main material components and production events for realizing the system-level MIC properties (i.e., metabolism, hereditary information, and compartmentalization) of biological cellular organization. These properties are well recognized in the theoretical study of life (reviewed in Cornish-Bowden and Cárdenas 2020), major evolutionary transitions (Smith and Szathmáry1997), and the minimal chemical foundations of life (Buddingh' and Van Hest 2017; Fellermann et al. 2007; Gánti 1975, 2003).

We will use the systemic MIC model of biomolecular cellular organization for holistic system-evolutionary contextualization of RNA presence in cell biology. At the same time, we acknowledge the RNA world hypothesis which posits that cellular life was preceded by an era of molecular-level RNA "life" where RNA molecules were both chemical catalysts as well as carries of genetic information (Fine and Pearlman 2023; Gilbert 1986; Müller et al. 2022). This era came to an end when cells evolved and the information storage function was transferred to DNA and the catalytic function to protein enzymes, and RNA molecules became strongly connected to the mechanisms of cellular protein production. We will be using the RNA world hypothesis as an additional systemic reference for examining new practical viewpoint for the characterization of RNA molecules and their presence in cells. And as synthetic biology already studies adaptive evolutionary capabilities of molecular-level entities (Arnold 2017), we will go one step further and suggest, that RNA molecules of cellular life can even be seen as bona fide molecular-level counterparts of organismal entities with genetically determined individual-level phenotypes contributing to their chance to (re)produced and survive in living cells. This is not a mere play with perspectives, but instead, it can help us build a view where we can understand the real actors of the biomolecular realm, especially RNA. The extended view that we propose provides a more comprehensive grounds for the translation of the resulting, more integrating theoretical understanding to meet the practice.

## 1.2 Metabolism, Hereditary Information and Compartmentalization: A Conceptual System-Level Outline of Cellular Biomolecular Organization

The whole-cell model that we use as a framework for considering RNA presence in cells is an original model by Segerståhl and Segerståhl (2022) describing a core set of components and processes that are universally needed by all cells to maintain a material foundation for realizing the MIC properties of biological cellular organization (Fig. 1.1). Unlike many minimal models[1] of living cells, the MIC model specifically remains connected to the general current knowledge of molecular cell biology while at the same time it is connected to the theoretical study of life and living system formalisms (Cornish-Bowden and Cárdenas 2020; Gánti 1975, 2003; Varela et al. 1974). Instead of repeating biochemical detail, the MIC network model provides an abstracted view of higher-level modular organizational aspects of molecular cell biology. This required taking terminology to a more general level that what is typical when describing molecular networks of cell biology. The terminology that is used, provides a link between the viewpoints of the biomolecular research and the more general theoretical study of life and organismal entities. This intermediate level of abstraction describes how molecular cell biology realizes an organization for the cell as living entity.

An interesting feature of the minimal MIC network model is that it visualizes self-production both for *material closure,* which results from the cell producing the components from which it consists of, and for *catalytic closure,* based on the structural properties of the network. It can be defined as an *autocatalytic set* meaning that the model describes biologically relevant systemic organization of components that collectively are capable of maintaining and reproducing themselves and the network of catalytic interactions that they form (Kauffman 1993), and even as a so-called *reflexively autocatalytic food-dependent (RAF) set*, which means that the network constructs its own catalysts using only simple raw materials that it acquires as "food" (Hordijk and Steel 2015, 2018). In other words, the oval nodes in Fig. 1.1 represent material products that the cell synthesizes and uses for realizing its MIC properties, and *at the same time*, these nodes refer to the components that provide the collective catalytic contribution for the synthesis of these same substances that establish the biological cell as a particular kind of distinctly identifiable organizational arrangement of energy and matter in the natural world.

[1] Focused on describing a minimal set of necessary components, aiming for models that contain as few details as possible.

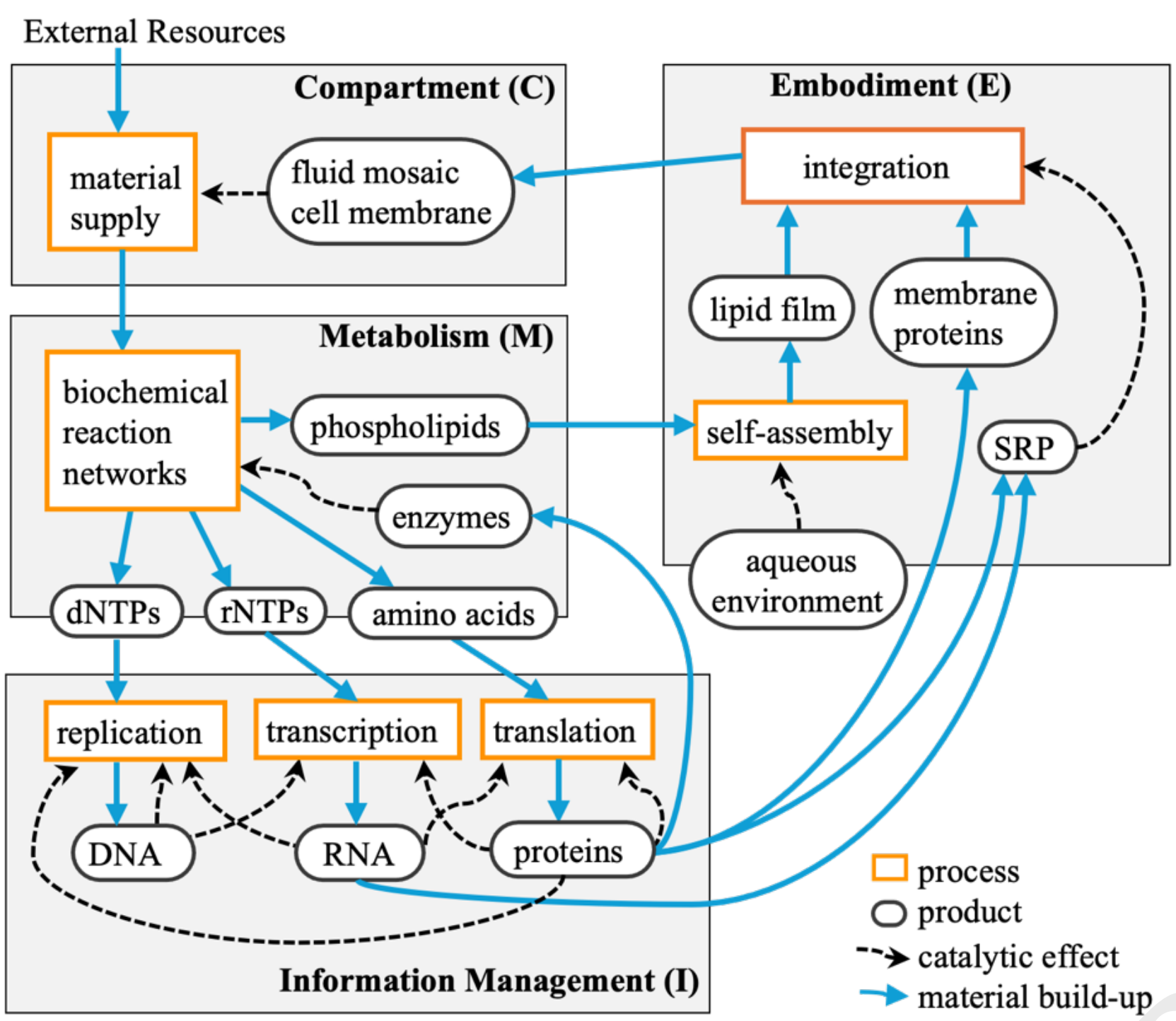

**Fig. 1.1** MIC network model of the biological cell. Minimal system view of universal organizational aspect of molecular cell biology featuring metabolism, information management, and compartmentalization (i.e., MIC) as the main system-level properties, that together with the encapsulation events produce embodiment and closure for an autocatalytic set of components and processes that constitute the cell as a holistic autopoietic living system entity. dNTPs, deoxyribonucleoside triphosphates; rNTPs, ribonucleoside triphosphates; SRP, signal recognition particle (modified from Segerståhl and Segerståhl 2022)

Cellular system-level compartmentalization is universally realized by being surrounded by a cell membrane. It mediates the flow of energy and matter between the cell and the environment including the acquisition of new raw materials for metabolic reactions. Protein enzymes are featured in the model as the material foundation on which cellular metabolism and the production of complex biomaterials is based on, including the synthesis of cell membrane lipids, nucleotide monomers of RNA and DNA strand synthesis, and amino acids of protein strand synthesis. The cell membrane is the most complex product type as it consists of chemical components that even individually are complex macromolecules (various kinds of lipids and proteins that can be modified or present as multimeric complexes). This complex level of material organization is generated by mechanisms that include signal recognition particle (SRP) mediated insertion of membrane proteins to the lipid phase, and self-organizing behavior of lipids in aqueous environments.

A notable feature of MIC network modularity is the conceptual separation of RNA, DNA, and protein production from the rest of cell metabolism. There are two main reasons for this. The synthesis of these three types of substances through template-based polymerization of their constituent of monomer units—ribonucleotides, deoxyribonucleotides and amino acids, respectively—into long linear information-bearing strands already makes these events and products unique among all other polymerization events and products of cell biology (e.g., long carbohydrates or fatty acids). In addition to this, while the mechanisms of RNA, DNA and protein synthesis are operationally autonomous, they form a collectively interdependent network of components and processes where the mechanisms of each process depend on having some components that are produced by another process of the network. The MIC model shows that this network exists as another, subsystemic RAF set inside the larger one that covers the whole cell. *This network will be at the center of our attention as we proceed with this article, and it is termed hereby as the RDP (i.e., RNA, DNA, protein) network of cell biology for easy referencing.*

## 1.3 The RDP Network—a Subsystemic Autocatalytic Set of RNA, DNA, and Protein Synthesis

A main incentive of the system-level study of molecular cell biology is to gain a more holistic understanding at the level of functional modules (Hartwell et al. 1999). The main outline of the subsystemic RDP network is the same in all living cells and would have been present in some basic form already in LUCA, together with a lipid membrane, a viral defense mechanism and a basis for cellular energy metabolism (Moody et al. 2024), although some of these perhaps only at the later stages of early cellular evolution (Jalasvuori and Bamford, 2008). Nonetheless, we will be using the RDP network as a distinct reference module against which RNA-related aspects of molecular cell biology can be reflected in an orderly manner. Fig. 1.2 provides a more focused view of it.

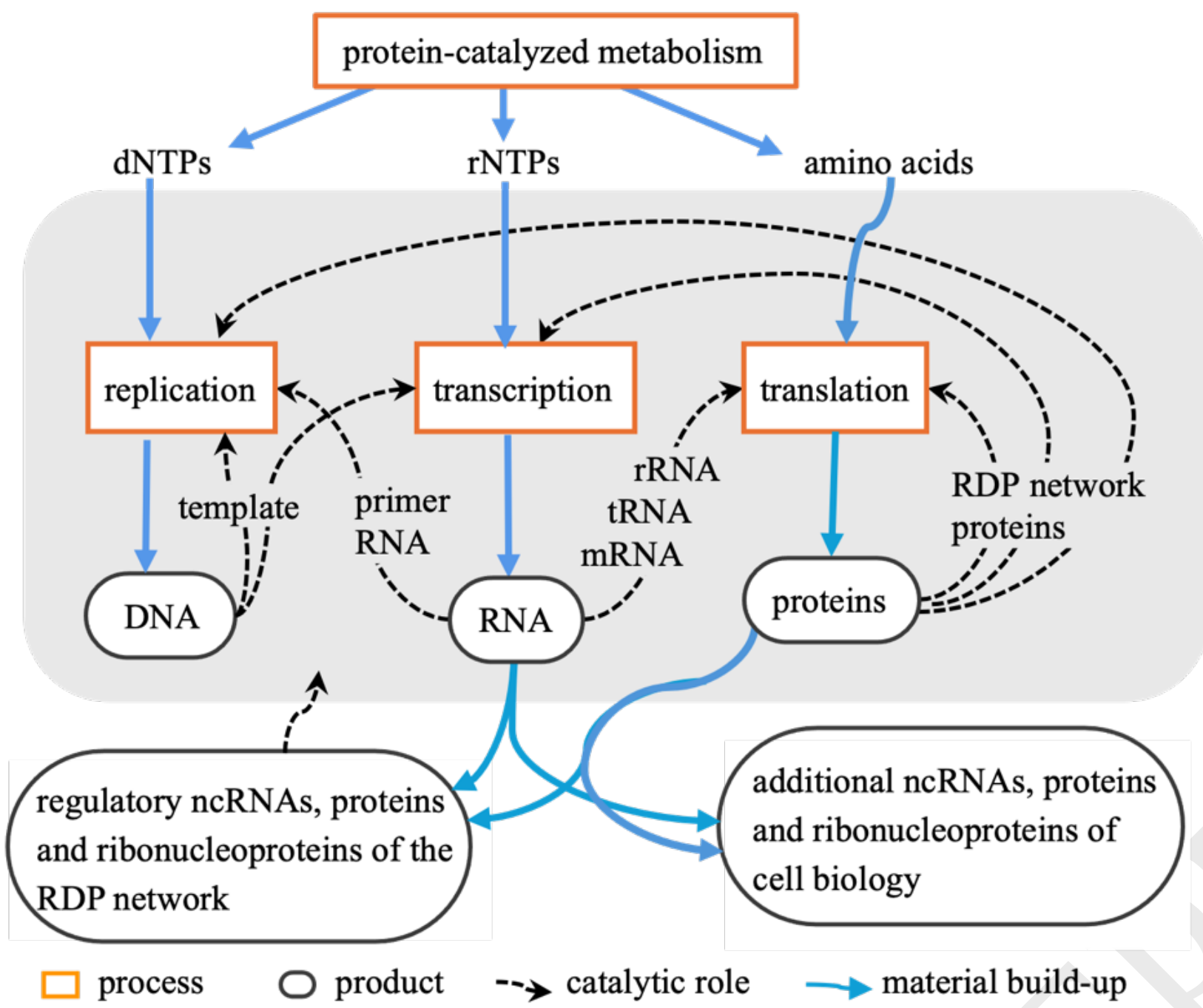

**Fig. 1.2** The RDP network of molecular cell biology. A schematic representation of the autocatalytic (RAF) set arrangement of components and processes constituting the biosynthesis network for the production of new RNA, DNA and protein (i.e., RDP) strands in living cells (gray background). The monomer units of RNA, DNA, and protein strands—ribonucleoside triphosphates (rNTPs), deoxyribonucleoside triphosphates (dNTPs), and amino acids—are considered to represent raw materials ("food")

The process events of the network are transcriptional RNA synthesis, polymerization of new DNA by the mechanisms of semiconservative DNA replication, and ribosomal synthesis of protein strands. The autocatalytic (RAF) network structure arises from the following connections: (1) DNA strands serve as templates both for the synthesis of new DNA as well as RNA strands, (2) the complex machinery of ribosomal protein synthesis requires cooperative contribution from mRNA, rRNA and tRNA molecules, and (3) proteins contribute significantly to the production of RNA, DNA, and protein strands. Additionally, the basic structure of the RDP network acknowledges the contribution of RNA primers to the mechanisms of semiconservative DNA replication—all cells need them to initiate DNA replication both on the leading strand and on every Okazaki fragment of the lagging strand. In all this, we turn our attention specifically to the constituent RNAs of this network. It is typical to use the functional perspective, to describe what the RNAs "deliver", as above, but to widen the scope we provide an additional schematic survey from an individual-level viewpoint of molecular phenotypes. This will be done by analyzing them against a simple formalism of general RNA shapes. This is motivated by the general notion that organismal phenotypes usually contain both reproduction and survival related characteristics, and justified by molecular cell biology being a realm where chemistry and biology overlap and structural characteristics of biomolecules can be approached from either perspective (Arnold 2017). This will enable making comparisons from a holistic system-evolutionary perspective of RDP network organization and dynamics between current existence of the RDP network RNAs, that are locked in the defining core of the RDP network interaction dynamics, and the viewpoints of the RNA world hypothesis regarding the events, selection pressures, and consequences of RNA becoming connected to the DNA–protein synthesis axis of protein coding gene expression of cellular life.

### 1.4 Holistic System-Level View of RNA Presence in Living Cells

Although focusing on a high level of abstraction, the RDP network describes a specific circuit module of molecular cell biology and remains connected to the conventional description paradigms of cell biology and biochemistry. Various types of relationships that exist in cells, both within and between the organizational levels of biopolymers, the RDP network, and the whole cell, can be examined from an RNA centered holistic systemic perspective.

### *1.4.1. Organismal Phenotypes of RNA Entities*

The four constituent RNA types of the RDP network are the mRNA, tRNA and rRNA molecules of protein coding gene expression, and the primer-RNA of semiconservative DNA replication. They all represent the same chemical substance, yet their individual-level structural properties are distinctly different. To provide new viewpoints for the study of RNA presence in cells, we adopt a systemic viewpoint that was originally used for abstract modeling of cells and multicellular organism by Segerståhl (2013). It considers that the phenotypic properties of any kind of organismal entities can be viewed and examined in terms of how they combine reproduction (R) versus survival (S) related phenotypic characteristics that often require conceptually opposite types of realizations. This viewpoint is connected to the two main components of Darwinian fitness, and we apply it to the characterization of molecular-level RNA entities in the following way: The degree to which an entity combines reproduction and survival related characteristics is considered as a phenotypic trait that is termed hereby as the *R/S ratio*. Linear single-stranded (ss) RNA conformation with the bases fully exposed and available for chemical interactions, corresponds to a conceptually open phenotype and having a high R/S ratio. A conceptually closed double-stranded (ds) RNA phenotype where complementary base-pairing provides physical protection for sequence identity, corresponds to having a low R/S ratio. However, the chemical structure of RNA strands is such that they tend to fold into conceptually intermediate conformations showing different kinds of combinations of ds-stalks and ss-loops. These structurally more complex phenotypes are considered to be conceptually intermediate in the light of the R/S formalism, providing a more balanced contribution to the two components of the R/S ratio. Based on this, a theoretical system axis as visualized in Fig. 1.3 can be formed and used as a reference standard against which R/S phenotypes of real biological RNAs can be compared in an orderly manner.

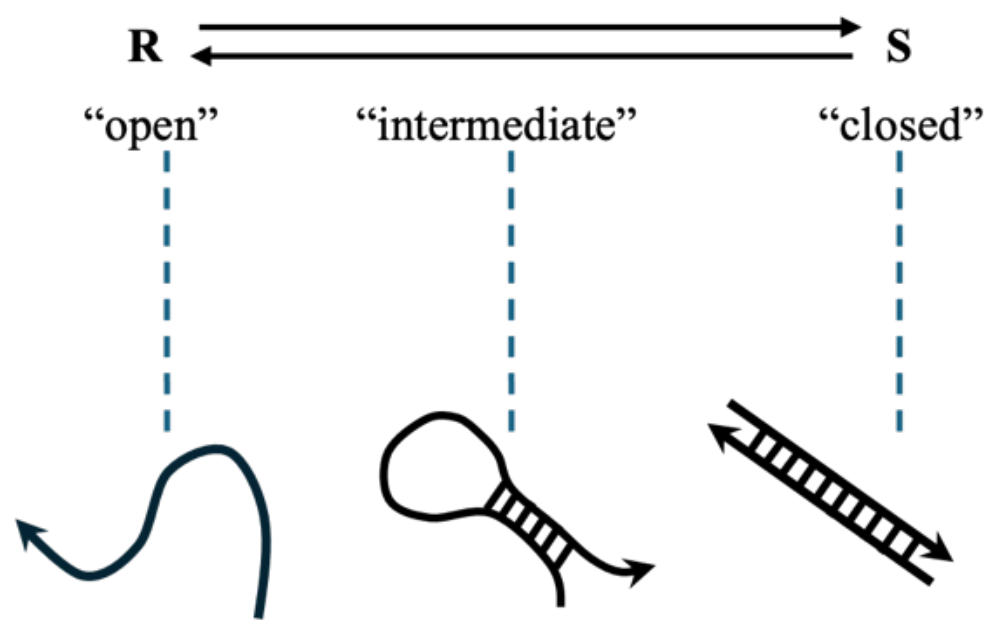

**Fig. 1.3** Schematic reference framework of organismal RNA phenotypes based on the ratio of reproduction (R) versus survival (S) related features in the expressed molecular-level folding pattern and how much it protects the sequence ("open" vs. "closed")

## *1.4.2. RNA Molecules of the RDP Network*

Based on what is generally known of the four classes of RDP network RNAs, we can observe that they perform their functional roles using very different kinds of naturally evolved, highly specialized R/S phenotypes. The RNA primers of semiconservative DNA replication are short-lived structures that do not exhibit any individual-level folding according to their sequence to perform their part. Instead, their presence creates functionally important RNA-DNA hybrid locations on chromosomal DNA, and the sequence patterns of primer-RNA only need to mirror that of the DNA templates accurately enough to remain in place until removed by cellular machinery. The physiological expressed phenotype of primer-RNA strands could be deemed as being linear from the individual perspective (high R/S ratio), yet at the same time being conceptually closed and protected, because while bound to the DNA, the RNA bases are chemically inaccessible and thus unavailable for chemical interactions.

In stark contrast to primer-RNA, the other three RNA types of the network express specialized individual-level phenotypes making use of diverse possibilities of RNA folding. The individual-level identity of mRNAs is above all based on the long single-strand sequence patterns of their coding regions. However, real biological mRNA contains also some sequence-dependent secondary structures in the untranslated regions, sometimes even in the coding region (Lewis et al. 2017). These are important for their ability to function properly with the rest of the components of the protein synthesizing genetic translation machinery. Intron splicing is an additional dynamic aspect of conformational mRNA complexity. And while the coding region is eventually only truly functioning when exposing its ss (while being managed by the ribosomal subunits) it can be protected by mRNA binding proteins from the moment it is transcribed until the moment it is being translated, during which time it effectively exists in a protected and non-operational, conceptually closed hybrid (ribonucleoprotein) state. Considering all these higher-level organizational aspects we conclude that real biological mRNA phenotypes are structurally plastic and dynamic. Based on the linear single-strand sequence of the coding regions being the most significant organizational level of the phenotype structure they can be defined as having, on an average, intermediate but R-biased R/S ratios.

tRNAs constitute the phenotypically most uniform class of RDP network RNAs. The prototypic sequence-based tRNA phenotype is a characteristic combination of conceptually open ss-loops and closed ds-stalks, and it is often used as a model example to demonstrate how RNA strands typically fold. Their folding happens efficiently via the formation of a 2D clover-leaf shape before the characteristic 3D L-shaped phenotype is adopted (e.g., Zhang 2024). In addition to the ones of the anticodon loop, there are also other key nucleotides in exact locations of the 1D ss-strand, that together define the functional precision of tRNA–ribosome interactions. Numerous post-transcriptional nucleotide modifications add further detail to the shape and the surface charge distribution patterns of functional tRNA molecules and their exact chemical interaction abilities.

And finally, the ribosomal RNA strands. These long well-structured RNAs express highly complex and precise individual-level 3D phenotypes while containing the oldest functional RNA motifs of cellular life (Bokov and Steinberg 2009). rRNAs have a sequence-based, higher-than-average tendency to form double-stranded segments but otherwise their grouping together is contextual (based on belonging to the ribosome) rather than based on structural or sequence similarities. Their ability to contribute to cell biology as core components of the RDP network stems from their sequence-based structural properties that enable them to assemble and act in concert with the ribosomal proteins and each other. The rRNA phenotypes are complex, higher-level organizational realizations of RNA folding, with ds-stalks and ss-loops as lower-scale motifs that can interact over longer distances, forming even higher-level structural arrangements while being kept together by the single-strand stretches that extend between the motifs in the 3D ribosomal subunit space. Thus, with the additional support of nucleotide modifications and protein assistance the pattern formation events of rRNA folding reach into far higher scales of physical organizational complexity that what a conceptual treatment of rRNA sequence patterns merely in the light of the R/S formalism or molecular-level physical chemistry might suggest, and we are, for now, left to wonder how exactly these impressive molecules should be fitted against the observation framework of R/S phenotypes.

Together, these four different kinds of RNAs of the RDP network form the core foundation of RNA presence in living cells. Many aspects of these real biological RNA phenotypes, such as the effective linearity of mRNA or the high-dimensional folding of rRNAs in association with ribosomal proteins, are not revealed directly by the respective linear sequence information that is preserved in DNA genes and knowledge of the folding dynamics of isolated RNA strands (minimizing free energy for thermodynamic stability) (Herschlag et al. 2018; Schroeder et al. 2004). The resulting biologically relevant individual-level properties of these RNA molecules—size, shape, and the distribution of reactive sites on the resulting molecular surface topology— are the key to the type of physical and chemical interactions and organizational behavior that they can have in cells. All this complexity is most relevant not only directly to the concrete material production of new DNA and proteins in living cells, but also more profoundly to the system-level maintenance of RDP network integrity that all cells depend on for being holistic depend on for their holistic existence.

### *1.4.3 Non-Coding RNAs and the RDP Network*

It is currently known that the RNA products of the RDP network's biosynthetic activities include numerous different kinds of ncRNAs with varying lengths, 3D structures, and interaction properties. Especially the discovery of a multitude of lncRNAs with diverse dynamic roles in eukaryotic cell biology has been paradigm-shifting (Mattick et al. 2023; Patil et al. 2013). (l)ncRNAs can regulate and control ongoing gene expression events, cause metabolic, behavioral or phenotypic change, control multicellular development or life cycle progression of unicellular organisms, provide protection against viral infections (e.g., RNA interference and CRISPR–Cas systems), and serve as the template for telomere elongation in the eukaryotic telomerase ribonucleoprotein complex, only to mention some examples (see, e.g., Guil and Esteller 2015; Ma et al. 2022, for a wider view).

The RDP network conceptualization provides a new kind of possibility to formally *differentiate between RNAs that belong to the system defining core components of the RDP network, and other types of cellular RNAs*. It enables us to make a novel kind of statement using new terminology that, notably, most (l)ncRNAs perform their functions specifically by interacting with the components and processes of the RDP network, while not belonging to the defining core components of the network (see Fig. 1.2). But they can regulate and modulate the synthesis of new RNA, DNA, and proteins in many concrete ways using different kinds of mechanisms and, for instance, be involved in the insertion (or removal) of epigenetic markings or non-random genetic change to the chromosomal DNA influencing thereby gene expression patterns.

The RDP network also produces all proteins of the cell—the constituent core proteins of the RDP network, metabolic enzymes, structural proteins, and functional proteins including those that are involved in signal transduction and information processing at the whole-cell level. Many protein-instantiated signals that originate in different parts of the cell and travel inside the cell space, for instance, in response to extracellular stimuli, in a sense "return" to the level of the RDP network that initially synthesized the respective proteins to modulate and modify RDP-network dynamics, with the final steps being cooperatively completed with additional (l)ncRNA and protein regulators of the RDP network core components. The resulting effects on gene expression can be very precise and effective and may even significantly change the state of the whole cell (Lu and Lionnet 2021; Paget 2015). From a wider system-level perspective (l)ncRNAs thus contribute alongside proteins to the holistic ability of biological cells to lead adaptive evolutionary lives in various kinds of environments.

### *1.4.4 Unique Opportunity for RNA Presented by the Membrane Dimension*

The cell membrane is a complex organizational substance dimension of macromolecular arrangements providing significant stability for cellular evolution with current life being based on only three main types of solutions, *the eubacterial, archaeal, and eukaryotic cell membranes* (Koga 2011; Nicolson 2014; Zhukov and Popov 2023) But within each type the compositional and structural details are very malleable, allowing for functional adaptive variability and phenotypic plasticity. The exact selection of proteins and lipids that are present at a particular instance, and their spatial arrangement within the membrane are key parameters that cells adjust to maintain a properly functioning cell membrane as the system-level interface that connects them to the wider environment.

The membrane is a unique dimension for RNA in the cell-system. It generates an intracellular niche-like space for localized co-existence and proximity of chemical substances and the chemical reaction networks of enzyme-based biochemistry (Buddingh' and Van Hest 2017; Torday 2016). The shape, size, and stability of early (proto)cellular lipid membrane assemblies probably varied (Deamer 2016; Lombard et al. 2012) and permanent compartmentalization may have been a later adaptation coupled to the evolution of membrane bending proteins (Laiterä and Lehto 2009; Mulkidjanian et al. 2009). Stabilization of the cell interface was a significant enabling factor for long-term adaptive co-evolution of cell membrane compositions, cellular biochemical networks, and RDP network dynamics, as natural environments are seldom non-changing. An adaptive, plastic, well-functioning gene expression—cell membrane axis is a significant system-level feature by which cellular life, and the RNAs in it claim living space in nature, and the cell signaling and control aspects featured in Sect. 1.4.3 are part of this holistic information-based big picture of molecular cell biology.

The RDP network controls the membrane dimension also rather directly by synthesizing the metabolic enzymes of membrane lipid synthesis as well as the proteins of functional cell membranes. In addition to this, the predominant mechanism that all cells use to insert proteins to the lipid phase is managed by the ancient SPR ribonucleoprotein complex that includes an RNA type that has been present in cells since they first emerged (Harris et al. 2021; Rosenblad et al. 2009). We might even extrapolate that from a general system-evolutionary perspective there is no particular reason why the RDP network could not use its ability to control the cell membrane dimension to manage whole cells so that they become constituent units of even higher-level biological assemblies of complex organismal existence (colonies, multicellular development). Maybe the RDP network could be seen as some kind of built-in enabling factor behind the ability of cells to find ever new ways to exist in nature, if necessary, by evolving into *increasingly more complex types of realizations of their MIC system organization.* This type of integrating systemic view of the large-scale evolutionary dynamics of molecular cell biology could perhaps be taken under closer examination in the future using, for instance, the kind of practical abstractions that we are presenting in this article as an attempt to deal with the organizational evolutionary complexity of RNA presence in living cells.

The membrane aspect might be especially connected to the origin and evolution of the eukaryotic cell that even in its simplest form includes distinct additional membrane-based compartments. The nuclear envelope, endomembrane system, mitochondria, chloroplasts and other plastids, endosymbionts, and multicellularity are true examples of membrane-based higher-level modularity in the physical realization of cells and it is a defining cell-level characteristic of eukaryotic life (Dacks and Field 2007). Why such higher-level assemblies would stabilize (as has occurred multiple times during the natural evolution history of life) is another question and part of the ongoing issue of how evolutionary theory can explain why the lower-level units do not disrupt higher-level assemblies (Bourrat et al. 2022; Smith and Szathmáry 1997; West et al. 2015). What we do know, however, is that it is precisely in the eukaryotes that epigenetics and (l)ncRNAs have been found to have an especially significant and powerful presence as regulatory and control elements of the RDP network dynamics and gene expression (Duclos et al. 2019; Mattick et al. 2023).

## 1.5 Integrating Viewpoints

In the course of this article, we have been integrating multiple different viewpoints in search of a new kind of holistic system-level view of the type of relationship that exists between RNA and cell biology. This has, however, been rather implicit as it is more about the methodology than a main thing to be noticed but now, we provide a breakdown of the four main ones that we have used.

The first viewpoint has been an organismal molecular-level view of individual RNA entities that belong to the constituent core components of the highly evolved and organized RDP network. They have been described as R/S entities that rely on their individual-level evolvable phenotypes for their collective ability to contribute to protein synthesis in living cells, and thereby to their own cell-bound existence in nature. The second one is the conventional viewpoint of the central dogma (Crick 1970), that focuses on the molecular mechanisms that transfer genetic hereditary sequence information between RNA, DNA and protein molecules. This can be seen directly in Figs. 1.1 and 1.2 where the process nodes of RNA and protein synthesis are termed "transcription" and "translation", respectively, knowing very well that the correspondence of these terms to all that these system-nodes represent in these models might not be an exact match. Using them is, however, a practical way to maintain connection to the conventional prevailing description paradigm of molecular cell biology while searching for novel conceptual ways to describe further abstract organizational aspects of cell biology.

The third viewpoint is a system-organizational viewpoint that focuses on the RDP network being organized as an autocatalytic (RAF) set. Autocatalytic sets have been used to study the chemical origins of life and real biochemical networks of living cells (Hordijk and Steel 2015, 2018; Sousa et al. 2015; Xavier et al. 2020). However, the MIC model is, to our knowledge, the first model of real biological cells that clearly identifies the RDP network as a distinct organizational module of cell biological network circuitry by showing it is a RAF set. The fourth viewpoint is that of the holistic MIC model approach where the RDP network is one part in a larger network abstraction that focuses on the system-level analysis of higher-level organizational aspects of cell biology and the whole cell. For instance, in the MIC model all RNA contribution that supports the "translation" node events are grouped together and represented by one arrow to avoid any unnecessary detail in what was intended to be a minimal model of the whole cell.

When studying complex system phenomena, the choice of viewpoints that are going to be used is part of larger issue of how to find the right kind of questions to be asked, what kind of methods are selected, and how the results are planned to be reported, and ideally, the choice of viewpoints should be based on active contemplation of different alternatives and traceable informed decisions (e.g., Keating et al. 2001; Levins 2006). This could be something to focus on more carefully when studying complex issues of biological organization.

## 1.6 System-Evolutionary Contemplations

We have provided an RNA-centered system-level view of the main types of RNAs that are fundamental to all cellular life using new kinds of holistic viewpoints that have been provided especially by the MIC network model of biological cellular organization. This has been the first instance of applying the abstract viewpoints of this network model in a practical way to better understand the evolutionary and functional relationships that connect RNAs to the rest of molecular cell biology.

### *1.6.1 Distributed System for the Usage and Storage of RNA information*

From a holistic system-evolutionary perspective, the RDP network presents itself as a distributed system for the usage and storage of RNA information, providing a stable foundation for adaptive evolutionary existence of RNA entities in living cells. The RNA, DNA, and protein components of the network (and of the whole cell) provide three qualitatively different kinds of material interfaces for presenting structural patterns of the cell's genetic hereditary information to the physical environment. While proteins connect the network to the realm of enzyme-catalyzed chemical reactions and energy metabolism, alpha-helical dsDNA is a superior substance for the physical preservation of sequence information (in comparison to dsRNA) and it is normally never degraded by active cellular mechanisms. With the many kinds of additional replication-based DNA repair mechanisms the genetic information of chromosomal DNA forms a permanent record of the set of RNA molecules that inhabit the cell. From RNA perspective, chromosomal DNA has effectively been assigned to take on the task of protecting RNA sequence information (sensu R/S formalism, see Sect. 1.4.1) and every time that the protein components of the gene expression machinery access chromosomal DNA, they (re)produce RNA using the DNA record of RNA information as template.

With the "parental" role allocated to the much more stable chromosomal DNA that contains the template strands of cellular RNA molecules, the latter can be seen as a "sterile" class of evolvable entities that are reproduced by proteins and stored by DNA. Individual RNAs can be temporarily lost without losing their defining information from the system allowing RNA molecules to have, from an organismal perspective, finite life spans. RNA, DNA, and protein strands also use their sequence-based phenotypes to interact with each other in a concrete physical way, and their structures can coevolve based on having their defining information held together by chromosomal DNA. But as the RDP network is a chemical system this requires compartmentalization as otherwise the molecular components would diffuse away and the network would cease to exist.

### *1.6.2 Did RNA Suffer of Benefit from the Transition to Cellular Life?*

Evolutionary study has been treating "independence" for a long time as a key criterion of evolutionary "success", and the transition of RNA "life" from an era of "independent" RNA world existence to the type of presence that it has in currently living cells with DNA and proteins, is typically being referred to in a negative sense. The standard view that is implicitly present in the way in which the RNA world hypothesis discusses the end of the RNA world era, is that RNA suffered some kind of loss or defeat when it became part of cell biology. However, given the new kinds of viewpoints that we have developed in this article we are willing to challenge this view. RNA matter could have been all the time leading an active and adaptive coevolutionary molecular-level "life" in living cells together with DNA, proteins, and lipid membranes. Cooperation and altruism are currently widely recognized as potentially highly significant supporting factors of adaptive organismal evolution, based on amounting experimental and empirical evidence from natural examples (e.g., Pennisi 2009). In this scenario RNA is taking full advantage of the physical properties of multiple additional substance dimensions that can express its information-related properties, each in their own respective ways, resulting in a far more complex and evolutionarily efficient biomolecular living system organization than what RNA alone could have accomplished. In this scenario RNA exists ultimately by managing the life that is led by biological cells. If viewed from this perspective, the stabilization of the RDP network as a fundamental early aspect of cellular biomolecular organization would have been a major advantageous evolutionary event for (populations of) RNA"life", releasing RNA from the double-duty constraint of having to be at all times both the catalysts as well as the keepers of their own system-defining hereditary information at all times.

### *1.6.3 On Stability and Variability*

The main structure of the RDP network and the MIC organization of the cell have remained the same for billions of years in connection to remarkable evolutionary emergence, stabilization, and diversification of numerous different kinds of additional variable features that have enabled cells to evolve and adapt and eventually to occupy nearly all kinds of Earth environments imaginable. Cellular life has tolerated, and in some cases perhaps even been looking for possible new ways to diversify genomes in support of dynamic, explorative molecular cell biology that can extend life to ever new ecological realms and spatiotemporal resolution scales of biological organization. The total complexity of RNA presence in currently living cells includes the everlasting coexistence of cellular life with viruses and the nowadays well-recognized consequences of HGT to cellular life and how it may shape the divergence of genetic codes (Jalasvuori 2020). In the light of all this, there is a distinct possibility, that rather than maximizing the preservation of cellular genome as unchanged and "intact" as possible, the ultimate system-level evolutionary guideline of molecular cell biology might be the *preservation of RDP network integrity and functional integrity of the cell as a holistic MIC system entity*. Endosymbionts, plasmids, organelles, and HGT would fall relatively effortlessly into this bigger picture. These genomes might be small, but when active they manifest *yet another way in which RNA information can be connected to biological cellular life*. There are organelles that have extremely reduced genomes (e.g. the mitochondrion) and might not use semiconservative replication to replicate their remaining chromosomal DNA, but they hold on to having an RDP network in some form, not just reminding of their past a cell-level MIC system entities, but keeping them still partly organismal (Kelly 2021; Roger et al. 2017) as opposed to just being complex macromolecular assemblies.

The large-scale evolution of molecular cell biology has been very dynamic also from the energetic perspective, especially in connection to historical macroevolutionary transitions when entirely new kinds of cells and organisms have emerged (Judson 2017; Okie et al. 2016). The predominant forms of energy that cells preferentially use and harvest from the environment have changed and diversified many times in the course of the natural history of biological evolution, yet the basic structure of the RDP network and of the MIC system organization of molecular cell biology have remained the same, as if being biological invariant organizational constants that define cells as living system entities. All this could be seen as having provided further support for active evolutionary RNA presence in living cells. From RNA perspective, the amount of ongoing gene expression is basically the quantity that determines how active and reproductive an RNA presence there is on Earth. The cutoff for what type and amount of novelty is good, and what is bad, what level of foreign elements or alterations can be tolerated, ultimately comes down to the reproduction and survival ability (the R/S ratio) of cells, whether existing alone or as constituent components of higher-level organizational assemblies and entities. This ratio must be a highly complex variable as we must not forget that all viral RNA is also produced in cells—and virus evolution can exploit the full range of strategic alternatives in terms of how intensely viral genes are expressed and how much stress infection imposes on the cells and organisms (Jalasvuori and Koonin 2015).

### *1.6.4 Ribosomal RNA and a Possible General Selection Pressure*

We finish this article with a brief discussion of the ribosome. It can be approached as an ancient organizing center of RDP network cell biology (e.g., Fer et al. 2025). Ribosomal proteins and rRNAs are closely coupled both from the evolutionary perspective and from the cell biological viewpoint of realizing the RDP network events in cells (Reuveni et al. 2017; Root-Bernstein and Root-Bernstein 2019). The early evolution of ribosomal assemblies was probably strongly connected to the aspect of protein synthesis, as the rRNA ribozyme that catalyzes peptide bond formation in all cells has one of the most ancient and preserved RNA patterns of cellular life (Bokov and Steinberg 2009). But in turn, ribosomal proteins are needed alongside various RNA binding proteins to assist in the high-dimensional folding process of long rRNAs which begins as soon as the strands begin to form (Rodgers and Woodson 2021), and ribosomes possibly still exhibit complex organizational features that evolved initially to optimized their own autocatalytic self-assembly as ribonucleoproteins (Reuveni et al. 2017).

In the bigger picture of organizational RDP network evolution and in search of holistic new perspectives and system-level viewpoints, we are left to wonder if the tight ribosomal RNA–protein couplings could be evolutionarily connected to avoiding the formation of dsRNA. This is because cells and viruses have most likely existed side by side in nature as long as cells have existed (Jalasvuori and Bamford 2008, 2009) as even LUCA probably already had a viral defense mechanism (Moody et al. 2024), and a distinct unifying aspect of cellular viral defense mechanisms is that they use dsRNA to detect viral infection. Thus, it may have been advantageous for early RNAs of (proto)cellular life to be associated with proteins, thereby avoiding inherent formation of long double-stranded regions that could result from folding or binding with a complementary strand. Removing the risk that dsRNA is "domestic" would make it easier for cells to protect RDP network integrity and thereby themselves against genetic invaders. Somewhat interestingly, using dsDNA as the storage medium (possibly in early evolutionary stages in the form of RNA–DNA ds-hybrids), as opposed to relying on dsRNA (the formal alternative in an RNA-only scenario, see Sect. 1.4.1) could also be seen from the viewpoint of this same potential selection pressure of allocating the use of dsRNA to viral defense mechanisms. Perhaps even the distinct absence of RNA-dependent RNA synthesis from the repertoire of cell biological mechanisms could be linked to this same potential selection pressure as this process would produce dsRNA if the old and new strand were not separated. We know that the reason for this absence is not that such mechanisms could not exist, as for example RNA phages use RNA-dependent mechanisms of RNA synthesis (Jalasvuori and Koonin 2015).

## Conclusion

In this article, we have shown new ways to apply novel abstract models of biomolecular cellular organization to better understand evolutionary complexity of real molecular cell biology and how new kinds of system-level viewpoints can be connected to existing ones in new ways. This is part of a wider search for concrete practical ways to reach some kind of new more holistic theoretical understanding of the biomolecular foundations of complexity evolution of cells, and of the multitude of biological mechanisms that can give rise to phenotypic plasticity and generate diversity at different organizational levels of biological organismal life and evolution. There is certainly still a long way to go. However, the system-oriented re-examination of existing knowledge of molecular cell biology and RNA presence in living cells that we have now conducted, might shed some light on the question of why RNA presence in living cells would be a much more complex and diverse phenomenon than what was initially believed. Furthermore, it no longer seems entirely confusing to us that in the course of the long adaptive evolutionary history of biological life many kinds of non-random mechanisms have evolved for causing, as well as for dealing with, epigenetic changes of gene expression and nonrandom genetic alteration of chromosomal DNA sequences, generating phenotypic diversity effectively in an open-ended way and being a major force of biological evolutionary life behind the existence of all kinds of cells and organisms that ever have inhabited the Earth.